\documentclass[%
reprint,
amsmath,amssymb,
aps,
prl,
]{revtex4-1}

\usepackage{graphicx}% Include figure files
\usepackage{dcolumn}% Align table columns on decimal point
\usepackage{bm}% bold math
\usepackage{mathrsfs}
\usepackage{textcomp}
\usepackage{subfigure}
\usepackage{amstext}
\usepackage{amsmath}
\newcommand{\bra}[1]{\langle#1|}
\newcommand{\ket}[1]{|#1\rangle}
\newcommand{\proj}[1]{|#1\rangle\langle#1|}

\newcommand{\mean}[1]{\langle #1 \rangle}

\usepackage[colorlinks,linkcolor=blue,anchorcolor=blue,citecolor=blue]{hyperref}

\begin{document}

\title{Topologically protected measurement of orbital angular momentum of light}

\author{Junfan Zhu}
\affiliation{College of Physics, Sichuan University, Chengdu 610064, China}
\author{An Wang}
\affiliation{College of Physics, Sichuan University, Chengdu 610064, China}
\author{Yurong Liu}
\affiliation{College of Physics, Sichuan University, Chengdu 610064, China}
\author{Fuhua Gao}%
\email{gaofuhua@scu.edu.cn}
\affiliation{College of Physics, Sichuan University, Chengdu 610064, China}%
\author{Zhiyou Zhang}%
\email{zhangzhiyou@scu.edu.cn}
\affiliation{College of Physics, Sichuan University, Chengdu 610064, China}%

\date{\today}

\begin{abstract}
	We develop a weak measurement scheme for measuring orbital angular momentum (OAM) of light based on the global topology in wave function. We introduce the spin-orbit coupling to transform the measurement of OAM to the pre- and postselected measurement of polarization. The OAM number can be precisely and promptly recognized using single-shot detection without the need for spatial resolution. More significantly, the measurement results exhibit topological robustness under random phase perturbations. This scheme has the potential to be applied as a paradigm in the OAM-based optical computing, metrology and communication.
\end{abstract}

\maketitle

\par The discovery of the quantum Hall effect sparked the insight of studying physical phenomena from a topological perspective \cite{PhysRevLett.45.494,PhysRevLett.49.405}. More profound realm was reached with the advent of topological insulators \cite{PhysRevLett.95.226801,PhysRevLett.96.106802}. The robustness of the edge states is attributed to the nontrivial topological invariant inherent in the energy level structure \cite{PhysRevLett.71.3697,RevModPhys.83.1057,RevModPhys.82.3045}. By analogy with the solid-state electron system, topology was introduced into photonic crystals to enable a unidirectional and backscattering-immune propagation of electromagnetic fields \cite{PhysRevLett.100.013905,wang2009observation}. Since then, topological photonic crystals have not only been used to simulate the novel Hall effects in condensed matter \cite{hafezi2011robust,rechtsman2013photonic,doi:10.1126/science.aaq0327,shalaev2019robust}, but also prompted some intriguing applications including topological lasers \cite{doi:10.1126/science.aao4551,zeng2020electrically}, topological light funnels \cite{doi:10.1126/science.aaz8727} and topological antennas \cite{lumer2020topological}.

\par Topology can also be manifested in the wave function of photon in free space. For a light field paraxially propagating along the $z$-direction, it permits a vortex phase profile that winds around a certain point. Considering a cylindrical coordinate $(r,\theta,z)$ and the presence of rotational symmetry, the wave function can be of the form $\psi(r,\theta)=\rho(r)\exp(i\ell\theta)$ with $\rho(r)$ denoting the amplitude and $\ell\theta$ the phase \cite{PhysRevA.108.040101}. The periodicity of the wave function implies the identity $\psi(r,\theta+2\pi)=\psi(r,\theta)$, which thereby requires that $\ell$ should be an integer. Furthermore, let $\hat{L}_z$ denote the $z$-component of the orbital angular momentum (OAM) operator and write it as $\hat{L}_z=-i\partial/\partial\theta$ in the coordinate representation \cite{BLIOKH20151}. The eigenvalue equation $\hat{L}_z\psi(r,\theta)=\ell\psi(r,\theta)$ indicates that $\ell$ characterizes the quantum number of OAM. Since the wave functions with different OAM numbers are orthogonal to each other, large-dimensional information can be encoded on single photons, which facilitates a broad range of applications in classical and quantum information \cite{10.1063/5.0054885,luo2015quantum,yang2022topological}. 

\par Accordingly, how to realize a precise and rapid measurement of OAM is an essential problem of wide concern. An orthodox method rests on the conversion from OAM beam to Gaussian beam, and then a spatial filter isolates the Gaussian beam to detect \cite{mair2001entanglement,Schlederer_2016}. The low efficiency comes from the limited utilization of photons, and $N$ possible OAM states  necessitating $N$ measurements. Other methods, such as interference \cite{Huang:13,Sztul:06,Zhao:20,PhysRevLett.112.153601} and diffraction \cite{Dai:15,Kotlyar:17,PhysRevLett.105.153601,Lavery:12,PhysRevLett.105.053904}, enable an indirect measurement in the far-field intensity distribution. The requirement of a detector with spatial resolution implies a restricted temporal response. More critically, the aforementioned methodologies are proposed based on light propagation without scattering or turbulence. It has been revealed that when an OAM beam is affected by atmospheric turbulence, the OAM mode will uniformly spread out \cite{PhysRevLett.94.153901,Tyler:09,Rodenburg:12,Malik:12,Ren:13}. In this case, the rotational symmetry is broken and the wave function is in the general form $\psi(r,\theta)=\rho(r,\theta)\exp[i\phi(r,\theta)]$, with $\phi(r,\theta)$ denoting the phase. The average OAM number is then given by
\begin{align}
	\overline{\ell}=\frac{1}{2\pi}\oint_C\vec{\nabla}\phi(r,\theta)\cdot d\vec{s}, \label{eq-ell}
\end{align}
where $C$ denotes a closed loop without self-intersection and $\vec{s}$ is an in-plane vector. Equation.~(\ref{eq-ell}) suggests a topological robustness that the average OAM number would remain unchanged even though there is spatial variation in wave function.

\par In this Letter, we are motivated by the topological protection to propose a scheme for measuring OAM. We employ the spin-orbit coupling to build a weak measurement system \cite{PhysRevLett.60.1351,Hosten787,PhysRevLett.102.173601,lundeen2011direct,kocsis2011observing,PhysRevLett.111.033604}, which permits single-shot detection without the need for spatial resolution. Considerable advantages are manifested in our scheme. The OAM number can be precisely and promptly measured. More significantly, we show that the measurement results can be robust under random phase perturbations. 

\begin{figure}
	\centering\includegraphics[scale=0.42]{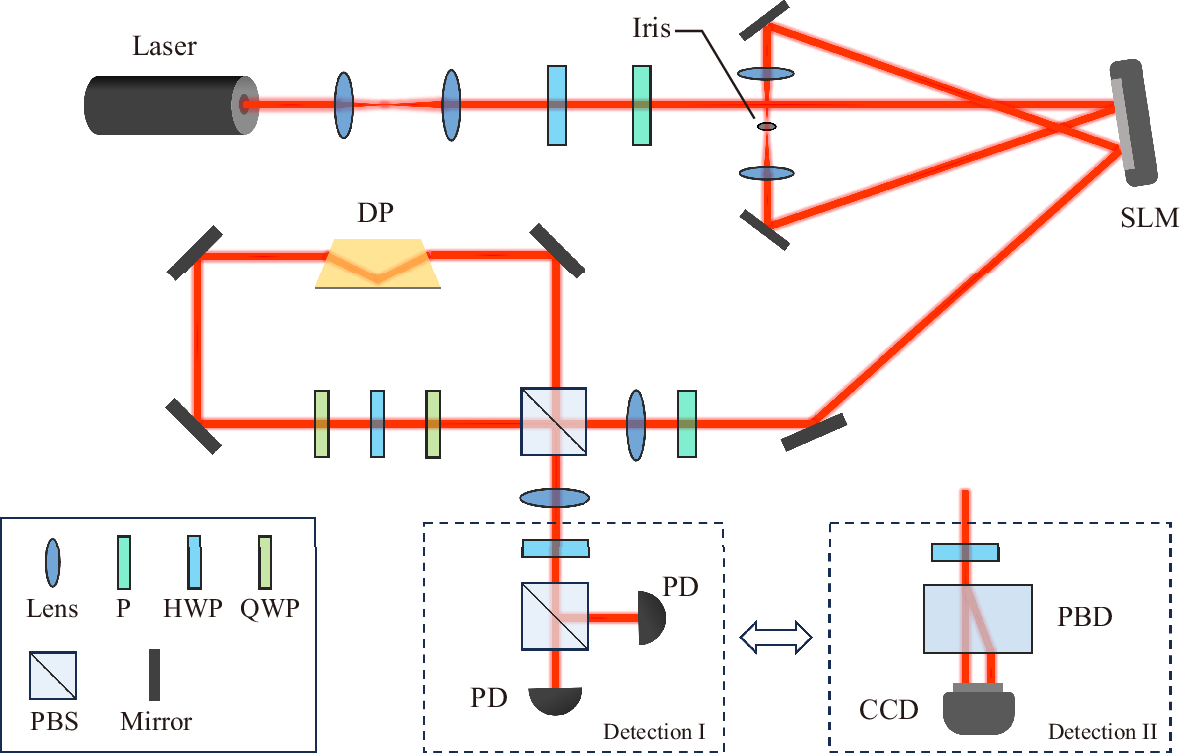}
	\caption{Schematic diagram of the experimental setup. P, polarizer; HWP, half wave plate; QWP, quarter wave plate; PBS, polarizing beam splitter; SLM, spatial light modulator; DP, Dove prism; PBD, polarizing beam displacer; PD, photon detector; CCD, charge couple device camera. Detection I suffices to measurement. We follow the optical configuration in Detection II for monitoring beam quality. See text for details.}
\end{figure}

\par The schematic diagram of our experimental setup is illustrated in Fig. 1. We use a He-Ne laser to emit a Gaussian beam with the wavelength of 633 nm, which is then expanded and collimated by a two-lens telescope to yield a waist of 0.66 mm. A half wave plate (HWP) and a polarizer are combined to adjust the light intensity and lock the polarization direction in the horizontal to match the working axis of the spatial light modulator (SLM). The display screen of SLM with the pixel pitch of 8 \textmu m is divided into left and right parts. We let the laser beam firstly impinge on the left part, where a fork-shaped grating is loaded to transform a Gaussian beam to an array of OAM beams \cite{doi:10.1080/09500340.2010.543292}. Two lenses and an iris make up a spatial filter such that only the first diffraction order is allowed to pass. The filtered OAM beam is redirected to impinge on the right part of SLM, where a time-varying random phase distribution is programmed. Each phase unit is composed of 8$\times$8 SLM pixels. After diffracting at a distance about 200 mm, the disturbed OAM wave function $\psi(r,\theta)$ is generated.

\par In the weak measurement system, a polarizer is placed with the polarizing direction set to $45^\circ$. Two lenses with the focal length of 250 mm establish a 4$f$ imaging system to remove the adverse effect of optical diffraction during the following measurement. We apply a polarizing beam splitter (PBS) to construct a rectangular Sagnac interferometer, wherein the common-path structure in propagating can make measurements robust to external disturbances, such as vibrations and thermal air currents. Through the PBS, the horizontally polarized component is transmitted while the vertically polarized component is reflected. The combination of two quarter wave plates (the fast axes set to $45^\circ$) and a HWP (the fast axis set to $67.5^\circ$) can introduce a Pancharatnam-Berry (PB) phase of $\pm\pi/4$ to the horizontal and the vertical polarization states, which are denoted by $\ket{H}$ and $\ket{V}$, respectively \cite{PhysRevA.80.012113}. The preselected state before the weak coupling is therefore given by
\begin{align}
	\ket{i}=\frac{1}{\sqrt{2}}(e^{i\frac{\pi}{4}}\ket{H}+\ket{V}).
\end{align}
The Dove prism (DP) is rotated by an angle of $\gamma/2$, such that the wave functions of two orthogonally polarized beams are rotated by an amount of $\pm\gamma$ in opposite directions, resulting in the spin-orbit coupling \cite{PhysRevLett.112.200401}. Assuming that $\gamma$ is sufficiently small, we can use a unitary operator to describe the weak coupling and expand it up to the first order, which is written as
\begin{align}
	U=e^{-i\gamma\hat{\sigma}\hat{L}_z}\approx1-i\gamma\hat{\sigma}\hat{L}_z,
\end{align}
where the Pauli spin operator is defined by $\hat{\sigma}=\proj{H}-\proj{V}$. The two beams are recombined by emerging from the PBS. We place a HWP (the fast axis set to $22.5^{\circ}$) and another PBS to project the photons in the state $(\ket{H}+\ket{V})/\sqrt{2}$ or $(\ket{H}-\ket{V})/\sqrt{2}$. In consideration of the PB phase induced to the state $\ket{V}$, two exit ports correspond to the postselected states
\begin{align}
	\ket{f_1}=\frac{1}{\sqrt{2}}(\ket{H}+e^{i\frac{\pi}{4}}\ket{V}),
\end{align}
and
\begin{align}
	\ket{f_2}=\frac{1}{\sqrt{2}}(\ket{H}-e^{i\frac{\pi}{4}}\ket{V}).
\end{align}
Consequently, the probability distributions are respectively given by
\begin{align}
	\Psi_1&(r,\theta)=|\bra{f_1}U\psi(r,\theta)\ket{i}|^2 \notag \\
	&=\frac{1}{2}\big\{\rho^2(r,\theta)\big[1+\gamma\partial_\theta\phi(r,\theta)\big]^2+\big[\gamma\partial_\theta\rho(r,\theta)\big]^2\big\},
\end{align}
and
\begin{align}
	\Psi_2&(r,\theta)=|\bra{f_2}U\psi(r,\theta)\ket{i}|^2 \notag \\
	&=\frac{1}{2}\big\{\rho^2(r,\theta)\big[1-\gamma\partial_\theta\phi(r,\theta)\big]^2+\big[\gamma\partial_\theta\rho(r,\theta)\big]^2\big\},
\end{align}
where $\partial_\theta$ is the abbreviation of $\partial/\partial\theta$. It is not difficult to note that $\Psi_1(r,\theta)-\Psi_2(r,\theta)=2\gamma\rho^2(r,\theta)\partial_\theta\phi(r,\theta)$, which implies that we may integrate the difference in probability to obtain the topologically invariant OAM number as suggested by Eq.~(\ref{eq-ell}).

\par In the presence of rotational symmetry, the wave function reads $\psi(r,\theta)=\rho(r)\exp(i\ell\theta)$. It follows that $\Psi_1(r,\theta)-\Psi_2(r,\theta)=2\gamma\ell\rho^2(r)$. Since a practical detector records results from a statistical accumulation of numerous photons, it would be convenient to define the dimensionless variable termed contrast ratio, which takes the form of
\begin{align}
	\mathcal{R}&=\frac{\mean{\Psi_1(r,\theta)-\Psi_2(r,\theta)}}{\mean{\Psi_1(r,\theta)+\Psi_2(r,\theta)}} \notag \\
	&=\frac{2\gamma\ell\mean{\rho^2(r)}}{\mean{\rho^2(r)}+\gamma^2\ell^2\mean{\rho^2(r)}} \notag \\
	&\approx2\gamma\ell, \label{eq-R}
\end{align}
where $\mean{g(r,\theta)}$ represents the integral of a function $g(r,\theta)$ over the entire space, i.e., $\iint_{-\infty}^{\infty}g(r,\theta)rdrd\theta$. In the third line of Eq.~(\ref{eq-R}), the higher-order terms with respect to $\gamma$ are neglected. Therefore, we can respectively place two photon detectors behind the two exit ports to measure the light intensities and thereafter calculate the contrast ratio. For the purpose of monitoring beam quality in real time, we can adopt a different strategy. A polarizing beam displacer that can separate the $\ket{H}$ state from the $\ket{V}$ state by 5 mm in space, and a charge coupled device (CCD) are employed.

\begin{figure}
	\centering\includegraphics[scale=0.44]{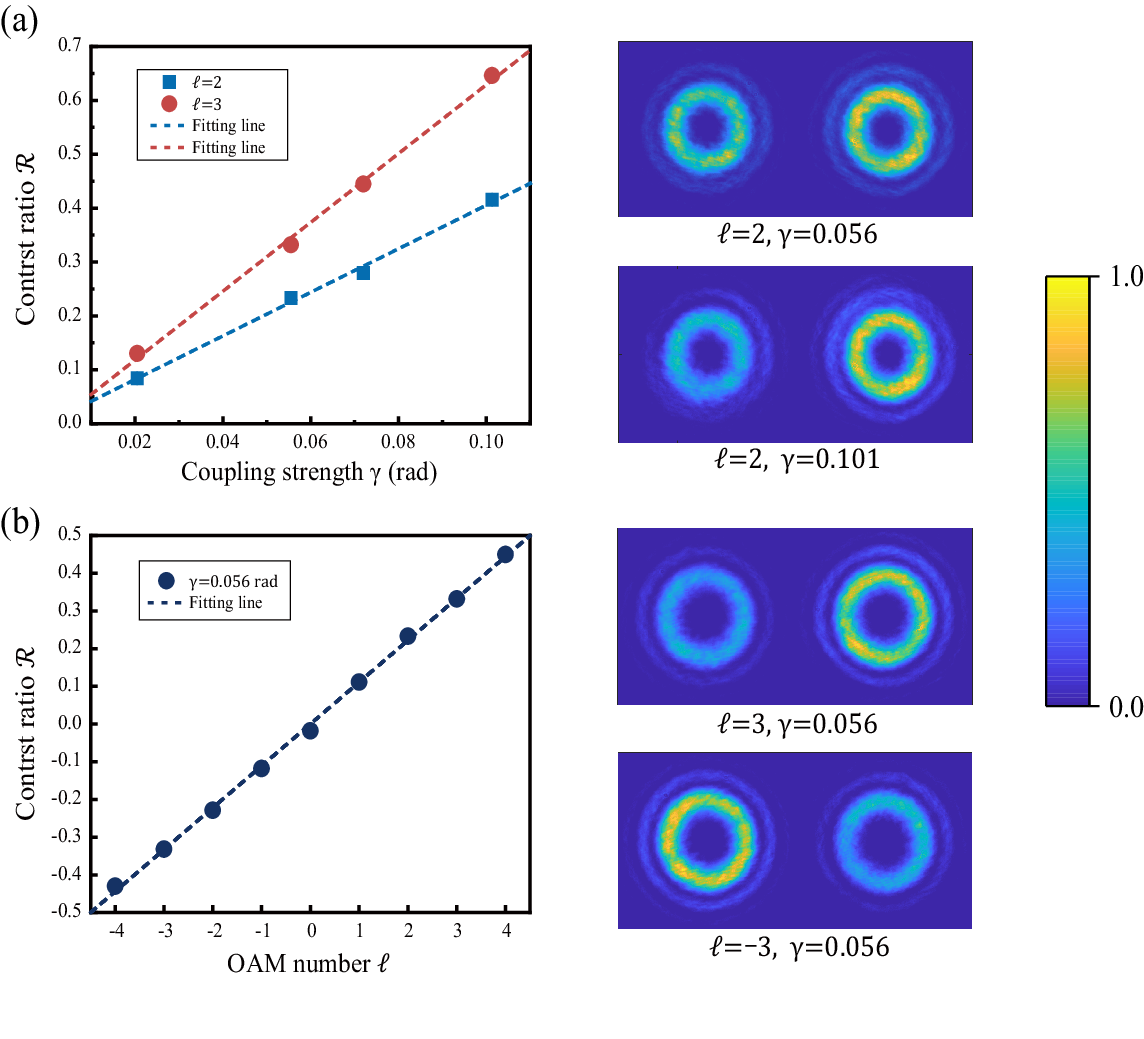}
	\caption{(a) Contrast ratio $\mathcal{R}$ for varying coupling strength $\gamma$. The blue and red correspond to the OAM number $\ell=2$ and $\ell=3$, respectively. The right shows the intensity distributions recorded by CCD when $\ell=2$ and $\gamma=0.056, 0.101$. (b) Contrast ratio $\mathcal{R}$ for varying OAM number $\ell$. The right shows the intensity distributions when $\gamma=0.056$ and $\ell=\pm3$.}
\end{figure}

\par For verification, first consider the case where no random phase is exerted, which means that the right part of SLM is plain. We use a micrometer to fine rotate the DP so as to slightly adjust the coupling strength. The left of Fig. 2(a) shows the contrast ratio $\mathcal{R}$ with respect to different coupling strengths ($\gamma$=0.021, 0.056, 0.072, and 0.101 rad). $\mathcal{R}$ is linearly proportional to $\gamma$ and the slope is greater for a larger OAM number. The right shows the normalized intensity distributions detected by CCD, giving a visualized contrast between $\Psi_1(r,\theta)$ and $\Psi_2(r,\theta)$ as $\gamma$ increasing. In Fig. 2(b), the values of $\mathcal{R}$ with $\gamma=0.056$ and $\ell$ ranging from $-4$ to 4 are measured. It indicates that the contrast ratio is a good pointer for OAM. The intensity distributions when $\ell=\pm3$ are also illustrated, which confirms that $\mean{\Psi_1(r,\theta)}$ is greater than $\mean{\Psi_2(r,\theta)}$ for a positive OAM number, whereas the relation is reversed as the sign of OAM is changed. Therefore, our scheme enables a flexible strategy in practical measurement. If the possible value of OAM is in a wide range, we should make the coupling strength small enough to include all the possibilities. If certain OAM numbers suffice to information transmission but the requirement for measurement accuracy is critical, we can increase the coupling strength to achieve a high contrast, instead.

\par In the case of a random phase disturbance, the rotational symmetry in wave function is broken. Assume that the randomness is not strong and the random phase can amount to a modulation $\Delta\varphi(r_s,\theta_s)$. The wave function $\psi(r,\theta)$ is derived by diffraction at a distance. Under this circumstance, we may decompose the amplitude into $\rho(r,\theta)=\rho(r)+\rho(r)K_1[\Delta\varphi(r_s,\theta_s)]$ and the phase into $\phi(r,\theta)=\ell\theta+K_2[\Delta\varphi(r_s,\theta_s)]$, with $K_1$ and $K_2$ denoting two linear operators. Disposing of the higher order terms of $\Delta\varphi(r_s,\theta_s)$ and making use of the topological invariance, the contrast ratio is thus given by
\begin{align}
	\mathcal{R}&=\frac{\mean{\Psi_1(r,\theta)-\Psi_2(r,\theta)}}{\mean{\Psi_1(r,\theta)+\Psi_2(r,\theta)}} \notag \\
	&=\frac{2\gamma\mean{\rho^2\partial_\theta\phi}}{\mean{\rho^2}+\gamma^2\mean{\rho^2(\partial_\theta\phi)^2}+\gamma^2\mean{(\partial_\theta\rho)^2}} \notag \\
	&\approx 2\gamma\ell-2\gamma^3\ell\big\langle\big\{[\partial_\theta K_1(\Delta\varphi)]^2+[\partial_\theta K_2(\Delta\varphi)]^2\big\}\rho^2\big\rangle, \label{eq-R2}
\end{align}
where we have neglected the dependency on coordinates for brevity. If there is no random phase, say $\Delta\varphi=0$, Eq.~(\ref{eq-R2}) will be reduced to Eq.~(\ref{eq-R}). As the value of $\Delta\varphi$ grows, $\mathcal{R}$ becomes smaller in that $[\partial_\theta K_1(\Delta\varphi)]^2$ and $[\partial_\theta K_2(\Delta\varphi)]^2$ are both non-negative. In a more quantitative way, we can define $\Delta\varphi_m$ to describe the degree of randomness, which means the phase modulation of each unit takes value randomly in the range between 0 and $\Delta\varphi_m$. The second term in the third line of Eq.~(\ref{eq-R2}) can be roughly considered of the order of $(\Delta\varphi_m)^2$. Consequently, $\mathcal{R}$ would be quadratically proportional to $\Delta\varphi_m$:
\begin{align}
	\mathcal{R}-2\gamma\ell \propto -2\gamma^3\ell(\Delta\varphi_m)^2. \label{eq-propto}
\end{align}
A threefold protection for OAM measurement is manifested in this relation. First, the deviation $\mathcal{R}-2\gamma\ell$ is cubically dependent on $\gamma$, indicating that a weaker coupling is better at suppressing the random phase disturbance (see Supplemental Material). The second protective effect comes from OAM itself. Light field with a lower OAM number can be more robust against disturbance. The quadratic dependence on $\Delta\varphi_m$ benefits from the topological invariance, showing that a near-zero randomness of disturbance can only lead to an insignificant degradation.

\begin{figure}
	\centering\includegraphics[scale=0.46]{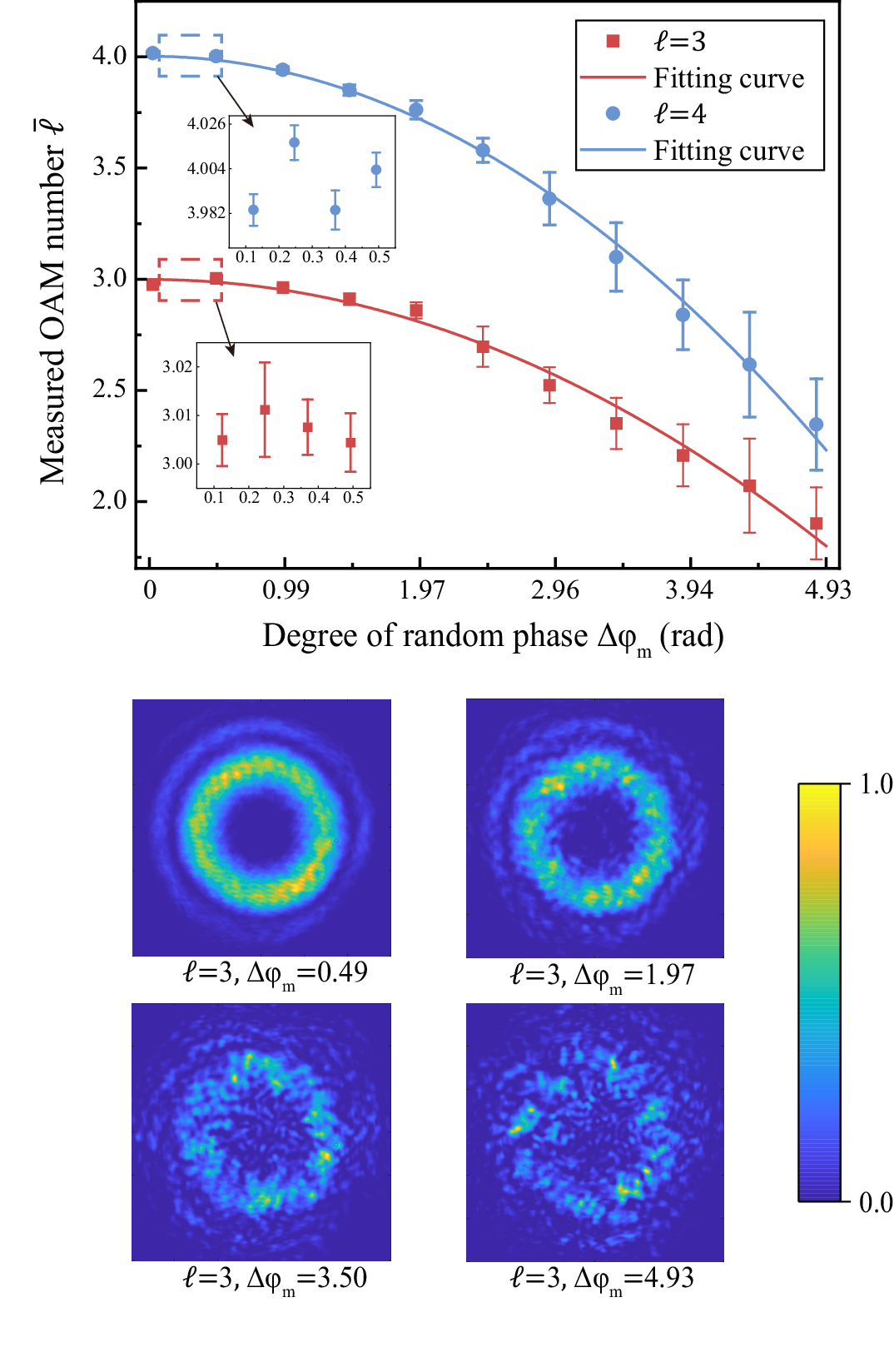}
	\caption{Measured OAM number $\overline{\ell}$ in a function of the degree of random phase $\Delta\varphi_m$. The red and blue correspond to $\ell=3$ and $\ell=4$, respectively. Details in the vicinity of $\Delta\varphi_m=0$ are shown in the insets. The bottom half gives the intensity distributions recorded by CCD for different $\Delta\varphi_m$.}
\end{figure}

\par In the experiment, we realize the phase disturbance by loading a random grayscale pattern onto the SLM. Each grayscale unit is associated with a random phase modulation between 0 and $\Delta\varphi_m$. We set the coupling strength to $\gamma=0.09$ rad and divide $\mathcal{R}$ by $2\gamma$ to indicate the measured OAM number. Respectively letting the OAM beams of $\ell=3$ and $\ell=4$ enter the weak measurement system, the measured $\overline{\ell}$ under different $\Delta\varphi_m$ are plotted in Fig. 3. It can be found that $\overline{\ell}$ is reduced as $\Delta\varphi_m$ grows, and a larger $\Delta\varphi_m$ corresponds to a higher reduction rate. The error bar is initially covered by data symbols when $\Delta\varphi_m$ is small, and later is lengthened with $\Delta\varphi_m$ increasing. The insets show the detail that the measured $\overline{\ell}$ is nearly constant in the vicinity of zero randomness. More specifically, the fitting curves for $\ell=3$ and $\ell=4$ are drawn based on the functions $\overline{\ell}_3=-0.048(\Delta\varphi_m)^2+2.993$ and $\overline{\ell}_4= -0.072(\Delta\varphi_m)^2+4.008$. The quadratic coefficient of $\overline{\ell}_4$ is greater than that of $\overline{\ell}_3$, which is in agreement with the prediction given by Eq.~(\ref{eq-propto}). The increasing of $\Delta\varphi_m$ results in a leakage of the measured $\overline{\ell}$ to a lower OAM number, i.e., crosstalk in measurement emerges. The bottom half of Fig. 3 provides a visual perception of beam quality under different degrees of randomness, which shows that the vortex structure would be hardly recognized when the randomness is severe. Nevertheless, the contrast ratio can partly tell the differences so long as an OAM is carried.

\par Finally, we have a further discussion of crosstalk. Channel capacity indicates the highest rate that can be achieved in information transmission \cite{6773024}. It is formally given by the maximum of the mutual information between the input and output of a channel \cite{cover1999elements}. In optical communication, if a photon can be in one of $N$ input states and its state can be measured without error in the output, the channel capacity reaches the theoretical maximum of $\log_2N$. The reduction of channel capacity suggests the emergence of crosstalk. 

\begin{figure}
	\centering\includegraphics[scale=0.42]{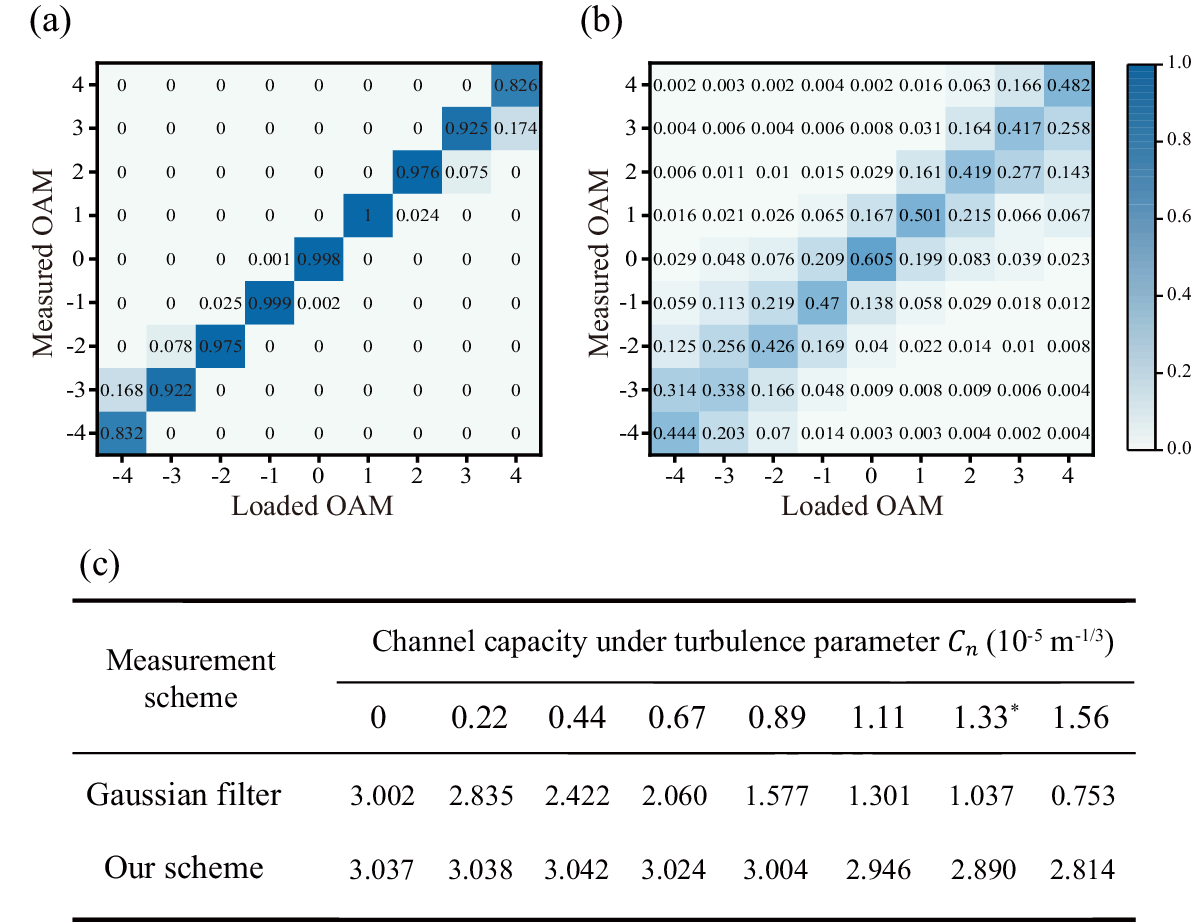}
	\caption{[(a) and (b)] Simulated confusion matrices for the turbulence parameter $C_n=1.33\times10^{-5}$ $\text{m}^{-1/3}$ and the OAM states from $\ell=-4$ to $\ell=4$, in the measurement scheme of (a) our scheme and (b) Gaussian filtering. (c) Comparison for channel capacity under different turbulence parameters between Gaussian filtering and our scheme.}
\end{figure}

\par By numerically simulating the transmission of the OAM states from $\ell=-4$ to $\ell=4$ under atmospheric turbulence \cite{Lukin:12}, we compare the channel capacity between the measurement scheme nominated as Gaussian filtering and the measurement scheme proposed by us. Gaussian filtering is widely used in the OAM detection, the principle of which is to transform an OAM beam back to a Gaussian beam by a conjugate phase modulation and then filter the Gaussian beam to detect its intensity \cite{mair2001entanglement,Schlederer_2016}. The parameter $C_n$ characterizes the strength of turbulence or, equivalently, the degree of randomness. We adjust the simulation parameters so that the two schemes have nearly the same channel capacity in the absence of turbulence. Figure. 4(a) and (b) present the confusion matrices respectively obtained in our scheme and Gaussian filtering when $C_n=1.33\times10^{-5}$ $\text{m}^{-1/3}$. Crosstalk is explicitly manifested in the confusion matrix. It can be noted that the measured OAM would leak evenly into the neighboring values for Gaussian filtering. While in our scheme, the contrast ratio only leads to a leakage towards a lower OAM number. The channel capacity with $C_n$ changing from 0 to $1.56\times10^{-5}$ $\text{m}^{-1/3}$ is tabulated in Fig. 4(c). The channel capacity dramatically decreases for Gaussian filtering. But in our scheme, the channel capacity is stable when $C_n$ is not too large. The topological protection is again manifested.

\par There is thus a trick if the contrast ratio is applied to recognize the OAM numbers. Since phase disturbance can always make the contrast ratio reduced, we may use the ceiling function to correct the OAM number when the crosstalk is not too severe. But this trick would be invalid in the case where a superposition state of OAM is input. Nevertheless, by combining with methods of demultiplexing \cite{PhysRevLett.88.257901}, our scheme is also possible to be applied in the sorting of OAM.

\par In conclusion, we have developed a weak measurement scheme to achieve a topologically protected measurement of OAM of light. We considered random phase distributions as disturbances impacted on the wave function. The quadratical dependence of the measured OAM on the degree of randomness showed a topological robustness. Moreover, single-shot detection without spatial resolution sufficed in our scheme, facilitating a precise and rapid measurement in practical. This work is promising to set a paradigm for OAM measurement in the applications of optical computing, metrology and communication.

\begin{acknowledgments}
	
	This work is supported by the Science Specialty Program of Sichuan University (Grant No.2020SCUNL210).
	
\end{acknowledgments}

\bibliography{ref}

\begin{thebibliography}{51}%
\makeatletter
\providecommand \@ifxundefined [1]{%
 \@ifx{#1\undefined}
}%
\providecommand \@ifnum [1]{%
 \ifnum #1\expandafter \@firstoftwo
 \else \expandafter \@secondoftwo
 \fi
}%
\providecommand \@ifx [1]{%
 \ifx #1\expandafter \@firstoftwo
 \else \expandafter \@secondoftwo
 \fi
}%
\providecommand \natexlab [1]{#1}%
\providecommand \enquote  [1]{``#1''}%
\providecommand \bibnamefont  [1]{#1}%
\providecommand \bibfnamefont [1]{#1}%
\providecommand \citenamefont [1]{#1}%
\providecommand \href@noop [0]{\@secondoftwo}%
\providecommand \href [0]{\begingroup \@sanitize@url \@href}%
\providecommand \@href[1]{\@@startlink{#1}\@@href}%
\providecommand \@@href[1]{\endgroup#1\@@endlink}%
\providecommand \@sanitize@url [0]{\catcode `\\12\catcode `\$12\catcode
  `\&12\catcode `\#12\catcode `\^12\catcode `\_12\catcode `\%12\relax}%
\providecommand \@@startlink[1]{}%
\providecommand \@@endlink[0]{}%
\providecommand \url  [0]{\begingroup\@sanitize@url \@url }%
\providecommand \@url [1]{\endgroup\@href {#1}{\urlprefix }}%
\providecommand \urlprefix  [0]{URL }%
\providecommand \Eprint [0]{\href }%
\providecommand \doibase [0]{http://dx.doi.org/}%
\providecommand \selectlanguage [0]{\@gobble}%
\providecommand \bibinfo  [0]{\@secondoftwo}%
\providecommand \bibfield  [0]{\@secondoftwo}%
\providecommand \translation [1]{[#1]}%
\providecommand \BibitemOpen [0]{}%
\providecommand \bibitemStop [0]{}%
\providecommand \bibitemNoStop [0]{.\EOS\space}%
\providecommand \EOS [0]{\spacefactor3000\relax}%
\providecommand \BibitemShut  [1]{\csname bibitem#1\endcsname}%
\let\auto@bib@innerbib\@empty
%</preamble>
\bibitem [{\citenamefont {Klitzing}\ \emph {et~al.}(1980)\citenamefont
  {Klitzing}, \citenamefont {Dorda},\ and\ \citenamefont
  {Pepper}}]{PhysRevLett.45.494}%
  \BibitemOpen
  \bibfield  {author} {\bibinfo {author} {\bibfnamefont {K.~v.}\ \bibnamefont
  {Klitzing}}, \bibinfo {author} {\bibfnamefont {G.}~\bibnamefont {Dorda}}, \
  and\ \bibinfo {author} {\bibfnamefont {M.}~\bibnamefont {Pepper}},\ }\href
  {\doibase 10.1103/PhysRevLett.45.494} {\bibfield  {journal} {\bibinfo
  {journal} {Phys. Rev. Lett.}\ }\textbf {\bibinfo {volume} {45}},\ \bibinfo
  {pages} {494} (\bibinfo {year} {1980})}\BibitemShut {NoStop}%
\bibitem [{\citenamefont {Thouless}\ \emph {et~al.}(1982)\citenamefont
  {Thouless}, \citenamefont {Kohmoto}, \citenamefont {Nightingale},\ and\
  \citenamefont {den Nijs}}]{PhysRevLett.49.405}%
  \BibitemOpen
  \bibfield  {author} {\bibinfo {author} {\bibfnamefont {D.~J.}\ \bibnamefont
  {Thouless}}, \bibinfo {author} {\bibfnamefont {M.}~\bibnamefont {Kohmoto}},
  \bibinfo {author} {\bibfnamefont {M.~P.}\ \bibnamefont {Nightingale}}, \ and\
  \bibinfo {author} {\bibfnamefont {M.}~\bibnamefont {den Nijs}},\ }\href
  {\doibase 10.1103/PhysRevLett.49.405} {\bibfield  {journal} {\bibinfo
  {journal} {Phys. Rev. Lett.}\ }\textbf {\bibinfo {volume} {49}},\ \bibinfo
  {pages} {405} (\bibinfo {year} {1982})}\BibitemShut {NoStop}%
\bibitem [{\citenamefont {Kane}\ and\ \citenamefont
  {Mele}(2005)}]{PhysRevLett.95.226801}%
  \BibitemOpen
  \bibfield  {author} {\bibinfo {author} {\bibfnamefont {C.~L.}\ \bibnamefont
  {Kane}}\ and\ \bibinfo {author} {\bibfnamefont {E.~J.}\ \bibnamefont
  {Mele}},\ }\href {\doibase 10.1103/PhysRevLett.95.226801} {\bibfield
  {journal} {\bibinfo  {journal} {Phys. Rev. Lett.}\ }\textbf {\bibinfo
  {volume} {95}},\ \bibinfo {pages} {226801} (\bibinfo {year}
  {2005})}\BibitemShut {NoStop}%
\bibitem [{\citenamefont {Bernevig}\ and\ \citenamefont
  {Zhang}(2006)}]{PhysRevLett.96.106802}%
  \BibitemOpen
  \bibfield  {author} {\bibinfo {author} {\bibfnamefont {B.~A.}\ \bibnamefont
  {Bernevig}}\ and\ \bibinfo {author} {\bibfnamefont {S.-C.}\ \bibnamefont
  {Zhang}},\ }\href {\doibase 10.1103/PhysRevLett.96.106802} {\bibfield
  {journal} {\bibinfo  {journal} {Phys. Rev. Lett.}\ }\textbf {\bibinfo
  {volume} {96}},\ \bibinfo {pages} {106802} (\bibinfo {year}
  {2006})}\BibitemShut {NoStop}%
\bibitem [{\citenamefont {Hatsugai}(1993)}]{PhysRevLett.71.3697}%
  \BibitemOpen
  \bibfield  {author} {\bibinfo {author} {\bibfnamefont {Y.}~\bibnamefont
  {Hatsugai}},\ }\href {\doibase 10.1103/PhysRevLett.71.3697} {\bibfield
  {journal} {\bibinfo  {journal} {Phys. Rev. Lett.}\ }\textbf {\bibinfo
  {volume} {71}},\ \bibinfo {pages} {3697} (\bibinfo {year}
  {1993})}\BibitemShut {NoStop}%
\bibitem [{\citenamefont {Qi}\ and\ \citenamefont
  {Zhang}(2011)}]{RevModPhys.83.1057}%
  \BibitemOpen
  \bibfield  {author} {\bibinfo {author} {\bibfnamefont {X.-L.}\ \bibnamefont
  {Qi}}\ and\ \bibinfo {author} {\bibfnamefont {S.-C.}\ \bibnamefont {Zhang}},\
  }\href {\doibase 10.1103/RevModPhys.83.1057} {\bibfield  {journal} {\bibinfo
  {journal} {Rev. Mod. Phys.}\ }\textbf {\bibinfo {volume} {83}},\ \bibinfo
  {pages} {1057} (\bibinfo {year} {2011})}\BibitemShut {NoStop}%
\bibitem [{\citenamefont {Hasan}\ and\ \citenamefont
  {Kane}(2010)}]{RevModPhys.82.3045}%
  \BibitemOpen
  \bibfield  {author} {\bibinfo {author} {\bibfnamefont {M.~Z.}\ \bibnamefont
  {Hasan}}\ and\ \bibinfo {author} {\bibfnamefont {C.~L.}\ \bibnamefont
  {Kane}},\ }\href {\doibase 10.1103/RevModPhys.82.3045} {\bibfield  {journal}
  {\bibinfo  {journal} {Rev. Mod. Phys.}\ }\textbf {\bibinfo {volume} {82}},\
  \bibinfo {pages} {3045} (\bibinfo {year} {2010})}\BibitemShut {NoStop}%
\bibitem [{\citenamefont {Wang}\ \emph {et~al.}(2008)\citenamefont {Wang},
  \citenamefont {Chong}, \citenamefont {Joannopoulos},\ and\ \citenamefont
  {Solja\ifmmode \check{c}\else \v{c}\fi{}i\ifmmode~\acute{c}\else
  \'{c}\fi{}}}]{PhysRevLett.100.013905}%
  \BibitemOpen
  \bibfield  {author} {\bibinfo {author} {\bibfnamefont {Z.}~\bibnamefont
  {Wang}}, \bibinfo {author} {\bibfnamefont {Y.~D.}\ \bibnamefont {Chong}},
  \bibinfo {author} {\bibfnamefont {J.~D.}\ \bibnamefont {Joannopoulos}}, \
  and\ \bibinfo {author} {\bibfnamefont {M.}~\bibnamefont {Solja\ifmmode
  \check{c}\else \v{c}\fi{}i\ifmmode~\acute{c}\else \'{c}\fi{}}},\ }\href
  {\doibase 10.1103/PhysRevLett.100.013905} {\bibfield  {journal} {\bibinfo
  {journal} {Phys. Rev. Lett.}\ }\textbf {\bibinfo {volume} {100}},\ \bibinfo
  {pages} {013905} (\bibinfo {year} {2008})}\BibitemShut {NoStop}%
\bibitem [{\citenamefont {Wang}\ \emph {et~al.}(2009)\citenamefont {Wang},
  \citenamefont {Chong}, \citenamefont {Joannopoulos},\ and\ \citenamefont
  {Solja{\v{c}}i{\'c}}}]{wang2009observation}%
  \BibitemOpen
  \bibfield  {author} {\bibinfo {author} {\bibfnamefont {Z.}~\bibnamefont
  {Wang}}, \bibinfo {author} {\bibfnamefont {Y.}~\bibnamefont {Chong}},
  \bibinfo {author} {\bibfnamefont {J.~D.}\ \bibnamefont {Joannopoulos}}, \
  and\ \bibinfo {author} {\bibfnamefont {M.}~\bibnamefont
  {Solja{\v{c}}i{\'c}}},\ }\href {\doibase 10.1038/nature08293} {\bibfield
  {journal} {\bibinfo  {journal} {Nature}\ }\textbf {\bibinfo {volume} {461}},\
  \bibinfo {pages} {772} (\bibinfo {year} {2009})}\BibitemShut {NoStop}%
\bibitem [{\citenamefont {Hafezi}\ \emph {et~al.}(2011)\citenamefont {Hafezi},
  \citenamefont {Demler}, \citenamefont {Lukin},\ and\ \citenamefont
  {Taylor}}]{hafezi2011robust}%
  \BibitemOpen
  \bibfield  {author} {\bibinfo {author} {\bibfnamefont {M.}~\bibnamefont
  {Hafezi}}, \bibinfo {author} {\bibfnamefont {E.~A.}\ \bibnamefont {Demler}},
  \bibinfo {author} {\bibfnamefont {M.~D.}\ \bibnamefont {Lukin}}, \ and\
  \bibinfo {author} {\bibfnamefont {J.~M.}\ \bibnamefont {Taylor}},\ }\href
  {\doibase 10.1038/nphys2063} {\bibfield  {journal} {\bibinfo  {journal}
  {Nature Physics}\ }\textbf {\bibinfo {volume} {7}},\ \bibinfo {pages} {907}
  (\bibinfo {year} {2011})}\BibitemShut {NoStop}%
\bibitem [{\citenamefont {Rechtsman}\ \emph {et~al.}(2013)\citenamefont
  {Rechtsman}, \citenamefont {Zeuner}, \citenamefont {Plotnik}, \citenamefont
  {Lumer}, \citenamefont {Podolsky}, \citenamefont {Dreisow}, \citenamefont
  {Nolte}, \citenamefont {Segev},\ and\ \citenamefont
  {Szameit}}]{rechtsman2013photonic}%
  \BibitemOpen
  \bibfield  {author} {\bibinfo {author} {\bibfnamefont {M.~C.}\ \bibnamefont
  {Rechtsman}}, \bibinfo {author} {\bibfnamefont {J.~M.}\ \bibnamefont
  {Zeuner}}, \bibinfo {author} {\bibfnamefont {Y.}~\bibnamefont {Plotnik}},
  \bibinfo {author} {\bibfnamefont {Y.}~\bibnamefont {Lumer}}, \bibinfo
  {author} {\bibfnamefont {D.}~\bibnamefont {Podolsky}}, \bibinfo {author}
  {\bibfnamefont {F.}~\bibnamefont {Dreisow}}, \bibinfo {author} {\bibfnamefont
  {S.}~\bibnamefont {Nolte}}, \bibinfo {author} {\bibfnamefont
  {M.}~\bibnamefont {Segev}}, \ and\ \bibinfo {author} {\bibfnamefont
  {A.}~\bibnamefont {Szameit}},\ }\href {\doibase 10.1038/nature12066}
  {\bibfield  {journal} {\bibinfo  {journal} {Nature}\ }\textbf {\bibinfo
  {volume} {496}},\ \bibinfo {pages} {196} (\bibinfo {year}
  {2013})}\BibitemShut {NoStop}%
\bibitem [{\citenamefont {Barik}\ \emph {et~al.}(2018)\citenamefont {Barik},
  \citenamefont {Karasahin}, \citenamefont {Flower}, \citenamefont {Cai},
  \citenamefont {Miyake}, \citenamefont {DeGottardi}, \citenamefont {Hafezi},\
  and\ \citenamefont {Waks}}]{doi:10.1126/science.aaq0327}%
  \BibitemOpen
  \bibfield  {author} {\bibinfo {author} {\bibfnamefont {S.}~\bibnamefont
  {Barik}}, \bibinfo {author} {\bibfnamefont {A.}~\bibnamefont {Karasahin}},
  \bibinfo {author} {\bibfnamefont {C.}~\bibnamefont {Flower}}, \bibinfo
  {author} {\bibfnamefont {T.}~\bibnamefont {Cai}}, \bibinfo {author}
  {\bibfnamefont {H.}~\bibnamefont {Miyake}}, \bibinfo {author} {\bibfnamefont
  {W.}~\bibnamefont {DeGottardi}}, \bibinfo {author} {\bibfnamefont
  {M.}~\bibnamefont {Hafezi}}, \ and\ \bibinfo {author} {\bibfnamefont
  {E.}~\bibnamefont {Waks}},\ }\href {\doibase 10.1126/science.aaq0327}
  {\bibfield  {journal} {\bibinfo  {journal} {Science}\ }\textbf {\bibinfo
  {volume} {359}},\ \bibinfo {pages} {666} (\bibinfo {year}
  {2018})}\BibitemShut {NoStop}%
\bibitem [{\citenamefont {Shalaev}\ \emph {et~al.}(2019)\citenamefont
  {Shalaev}, \citenamefont {Walasik}, \citenamefont {Tsukernik}, \citenamefont
  {Xu},\ and\ \citenamefont {Litchinitser}}]{shalaev2019robust}%
  \BibitemOpen
  \bibfield  {author} {\bibinfo {author} {\bibfnamefont {M.~I.}\ \bibnamefont
  {Shalaev}}, \bibinfo {author} {\bibfnamefont {W.}~\bibnamefont {Walasik}},
  \bibinfo {author} {\bibfnamefont {A.}~\bibnamefont {Tsukernik}}, \bibinfo
  {author} {\bibfnamefont {Y.}~\bibnamefont {Xu}}, \ and\ \bibinfo {author}
  {\bibfnamefont {N.~M.}\ \bibnamefont {Litchinitser}},\ }\href {\doibase
  10.1038/s41565-018-0297-6} {\bibfield  {journal} {\bibinfo  {journal} {Nature
  nanotechnology}\ }\textbf {\bibinfo {volume} {14}},\ \bibinfo {pages} {31}
  (\bibinfo {year} {2019})}\BibitemShut {NoStop}%
\bibitem [{\citenamefont {Bahari}\ \emph {et~al.}(2017)\citenamefont {Bahari},
  \citenamefont {Ndao}, \citenamefont {Vallini}, \citenamefont {Amili},
  \citenamefont {Fainman},\ and\ \citenamefont
  {Kanté}}]{doi:10.1126/science.aao4551}%
  \BibitemOpen
  \bibfield  {author} {\bibinfo {author} {\bibfnamefont {B.}~\bibnamefont
  {Bahari}}, \bibinfo {author} {\bibfnamefont {A.}~\bibnamefont {Ndao}},
  \bibinfo {author} {\bibfnamefont {F.}~\bibnamefont {Vallini}}, \bibinfo
  {author} {\bibfnamefont {A.~E.}\ \bibnamefont {Amili}}, \bibinfo {author}
  {\bibfnamefont {Y.}~\bibnamefont {Fainman}}, \ and\ \bibinfo {author}
  {\bibfnamefont {B.}~\bibnamefont {Kanté}},\ }\href {\doibase
  10.1126/science.aao4551} {\bibfield  {journal} {\bibinfo  {journal}
  {Science}\ }\textbf {\bibinfo {volume} {358}},\ \bibinfo {pages} {636}
  (\bibinfo {year} {2017})}\BibitemShut {NoStop}%
\bibitem [{\citenamefont {Zeng}\ \emph {et~al.}(2020)\citenamefont {Zeng},
  \citenamefont {Chattopadhyay}, \citenamefont {Zhu}, \citenamefont {Qiang},
  \citenamefont {Li}, \citenamefont {Jin}, \citenamefont {Li}, \citenamefont
  {Davies}, \citenamefont {Linfield}, \citenamefont {Zhang} \emph
  {et~al.}}]{zeng2020electrically}%
  \BibitemOpen
  \bibfield  {author} {\bibinfo {author} {\bibfnamefont {Y.}~\bibnamefont
  {Zeng}}, \bibinfo {author} {\bibfnamefont {U.}~\bibnamefont {Chattopadhyay}},
  \bibinfo {author} {\bibfnamefont {B.}~\bibnamefont {Zhu}}, \bibinfo {author}
  {\bibfnamefont {B.}~\bibnamefont {Qiang}}, \bibinfo {author} {\bibfnamefont
  {J.}~\bibnamefont {Li}}, \bibinfo {author} {\bibfnamefont {Y.}~\bibnamefont
  {Jin}}, \bibinfo {author} {\bibfnamefont {L.}~\bibnamefont {Li}}, \bibinfo
  {author} {\bibfnamefont {A.~G.}\ \bibnamefont {Davies}}, \bibinfo {author}
  {\bibfnamefont {E.~H.}\ \bibnamefont {Linfield}}, \bibinfo {author}
  {\bibfnamefont {B.}~\bibnamefont {Zhang}},  \emph {et~al.},\ }\href {\doibase
  10.1038/s41586-020-1981-x} {\bibfield  {journal} {\bibinfo  {journal}
  {Nature}\ }\textbf {\bibinfo {volume} {578}},\ \bibinfo {pages} {246}
  (\bibinfo {year} {2020})}\BibitemShut {NoStop}%
\bibitem [{\citenamefont {Weidemann}\ \emph {et~al.}(2020)\citenamefont
  {Weidemann}, \citenamefont {Kremer}, \citenamefont {Helbig}, \citenamefont
  {Hofmann}, \citenamefont {Stegmaier}, \citenamefont {Greiter}, \citenamefont
  {Thomale},\ and\ \citenamefont {Szameit}}]{doi:10.1126/science.aaz8727}%
  \BibitemOpen
  \bibfield  {author} {\bibinfo {author} {\bibfnamefont {S.}~\bibnamefont
  {Weidemann}}, \bibinfo {author} {\bibfnamefont {M.}~\bibnamefont {Kremer}},
  \bibinfo {author} {\bibfnamefont {T.}~\bibnamefont {Helbig}}, \bibinfo
  {author} {\bibfnamefont {T.}~\bibnamefont {Hofmann}}, \bibinfo {author}
  {\bibfnamefont {A.}~\bibnamefont {Stegmaier}}, \bibinfo {author}
  {\bibfnamefont {M.}~\bibnamefont {Greiter}}, \bibinfo {author} {\bibfnamefont
  {R.}~\bibnamefont {Thomale}}, \ and\ \bibinfo {author} {\bibfnamefont
  {A.}~\bibnamefont {Szameit}},\ }\href {\doibase 10.1126/science.aaz8727}
  {\bibfield  {journal} {\bibinfo  {journal} {Science}\ }\textbf {\bibinfo
  {volume} {368}},\ \bibinfo {pages} {311} (\bibinfo {year}
  {2020})}\BibitemShut {NoStop}%
\bibitem [{\citenamefont {Lumer}\ and\ \citenamefont
  {Engheta}(2020)}]{lumer2020topological}%
  \BibitemOpen
  \bibfield  {author} {\bibinfo {author} {\bibfnamefont {Y.}~\bibnamefont
  {Lumer}}\ and\ \bibinfo {author} {\bibfnamefont {N.}~\bibnamefont
  {Engheta}},\ }\href {\doibase 10.1021/acsphotonics.0c00797} {\bibfield
  {journal} {\bibinfo  {journal} {ACS Photonics}\ }\textbf {\bibinfo {volume}
  {7}},\ \bibinfo {pages} {2244} (\bibinfo {year} {2020})}\BibitemShut
  {NoStop}%
\bibitem [{\citenamefont {Jalali~Mehrabad}\ \emph {et~al.}(2023)\citenamefont
  {Jalali~Mehrabad}, \citenamefont {Mittal},\ and\ \citenamefont
  {Hafezi}}]{PhysRevA.108.040101}%
  \BibitemOpen
  \bibfield  {author} {\bibinfo {author} {\bibfnamefont {M.}~\bibnamefont
  {Jalali~Mehrabad}}, \bibinfo {author} {\bibfnamefont {S.}~\bibnamefont
  {Mittal}}, \ and\ \bibinfo {author} {\bibfnamefont {M.}~\bibnamefont
  {Hafezi}},\ }\href {\doibase 10.1103/PhysRevA.108.040101} {\bibfield
  {journal} {\bibinfo  {journal} {Phys. Rev. A}\ }\textbf {\bibinfo {volume}
  {108}},\ \bibinfo {pages} {040101} (\bibinfo {year} {2023})}\BibitemShut
  {NoStop}%
\bibitem [{\citenamefont {Bliokh}\ and\ \citenamefont
  {Nori}(2015)}]{BLIOKH20151}%
  \BibitemOpen
  \bibfield  {author} {\bibinfo {author} {\bibfnamefont {K.~Y.}\ \bibnamefont
  {Bliokh}}\ and\ \bibinfo {author} {\bibfnamefont {F.}~\bibnamefont {Nori}},\
  }\href {\doibase https://doi.org/10.1016/j.physrep.2015.06.003} {\bibfield
  {journal} {\bibinfo  {journal} {Physics Reports}\ }\textbf {\bibinfo {volume}
  {592}},\ \bibinfo {pages} {1} (\bibinfo {year} {2015})},\ \bibinfo {note}
  {transverse and longitudinal angular momenta of light}\BibitemShut {NoStop}%
\bibitem [{\citenamefont {Willner}\ \emph {et~al.}(2021)\citenamefont
  {Willner}, \citenamefont {Pang}, \citenamefont {Song}, \citenamefont {Zou},\
  and\ \citenamefont {Zhou}}]{10.1063/5.0054885}%
  \BibitemOpen
  \bibfield  {author} {\bibinfo {author} {\bibfnamefont {A.~E.}\ \bibnamefont
  {Willner}}, \bibinfo {author} {\bibfnamefont {K.}~\bibnamefont {Pang}},
  \bibinfo {author} {\bibfnamefont {H.}~\bibnamefont {Song}}, \bibinfo {author}
  {\bibfnamefont {K.}~\bibnamefont {Zou}}, \ and\ \bibinfo {author}
  {\bibfnamefont {H.}~\bibnamefont {Zhou}},\ }\href {\doibase
  10.1063/5.0054885} {\bibfield  {journal} {\bibinfo  {journal} {Applied
  Physics Reviews}\ }\textbf {\bibinfo {volume} {8}},\ \bibinfo {pages}
  {041312} (\bibinfo {year} {2021})}\BibitemShut {NoStop}%
\bibitem [{\citenamefont {Luo}\ \emph {et~al.}(2015)\citenamefont {Luo},
  \citenamefont {Zhou}, \citenamefont {Li}, \citenamefont {Xu}, \citenamefont
  {Guo},\ and\ \citenamefont {Zhou}}]{luo2015quantum}%
  \BibitemOpen
  \bibfield  {author} {\bibinfo {author} {\bibfnamefont {X.-W.}\ \bibnamefont
  {Luo}}, \bibinfo {author} {\bibfnamefont {X.}~\bibnamefont {Zhou}}, \bibinfo
  {author} {\bibfnamefont {C.-F.}\ \bibnamefont {Li}}, \bibinfo {author}
  {\bibfnamefont {J.-S.}\ \bibnamefont {Xu}}, \bibinfo {author} {\bibfnamefont
  {G.-C.}\ \bibnamefont {Guo}}, \ and\ \bibinfo {author} {\bibfnamefont
  {Z.-W.}\ \bibnamefont {Zhou}},\ }\href {\doibase 10.1038/ncomms8704}
  {\bibfield  {journal} {\bibinfo  {journal} {Nature communications}\ }\textbf
  {\bibinfo {volume} {6}},\ \bibinfo {pages} {7704} (\bibinfo {year}
  {2015})}\BibitemShut {NoStop}%
\bibitem [{\citenamefont {Yang}\ \emph {et~al.}(2022)\citenamefont {Yang},
  \citenamefont {Zhang}, \citenamefont {Liao}, \citenamefont {Liu},
  \citenamefont {Zhou}, \citenamefont {Zhou}, \citenamefont {Xu}, \citenamefont
  {Han}, \citenamefont {Li},\ and\ \citenamefont {Guo}}]{yang2022topological}%
  \BibitemOpen
  \bibfield  {author} {\bibinfo {author} {\bibfnamefont {M.}~\bibnamefont
  {Yang}}, \bibinfo {author} {\bibfnamefont {H.-Q.}\ \bibnamefont {Zhang}},
  \bibinfo {author} {\bibfnamefont {Y.-W.}\ \bibnamefont {Liao}}, \bibinfo
  {author} {\bibfnamefont {Z.-H.}\ \bibnamefont {Liu}}, \bibinfo {author}
  {\bibfnamefont {Z.-W.}\ \bibnamefont {Zhou}}, \bibinfo {author}
  {\bibfnamefont {X.-X.}\ \bibnamefont {Zhou}}, \bibinfo {author}
  {\bibfnamefont {J.-S.}\ \bibnamefont {Xu}}, \bibinfo {author} {\bibfnamefont
  {Y.-J.}\ \bibnamefont {Han}}, \bibinfo {author} {\bibfnamefont {C.-F.}\
  \bibnamefont {Li}}, \ and\ \bibinfo {author} {\bibfnamefont {G.-C.}\
  \bibnamefont {Guo}},\ }\href {\doibase 10.1038/s41467-022-29779-3} {\bibfield
   {journal} {\bibinfo  {journal} {Nature Communications}\ }\textbf {\bibinfo
  {volume} {13}},\ \bibinfo {pages} {2040} (\bibinfo {year}
  {2022})}\BibitemShut {NoStop}%
\bibitem [{\citenamefont {Mair}\ \emph {et~al.}(2001)\citenamefont {Mair},
  \citenamefont {Vaziri}, \citenamefont {Weihs},\ and\ \citenamefont
  {Zeilinger}}]{mair2001entanglement}%
  \BibitemOpen
  \bibfield  {author} {\bibinfo {author} {\bibfnamefont {A.}~\bibnamefont
  {Mair}}, \bibinfo {author} {\bibfnamefont {A.}~\bibnamefont {Vaziri}},
  \bibinfo {author} {\bibfnamefont {G.}~\bibnamefont {Weihs}}, \ and\ \bibinfo
  {author} {\bibfnamefont {A.}~\bibnamefont {Zeilinger}},\ }\href {\doibase
  10.1038/35085529} {\bibfield  {journal} {\bibinfo  {journal} {Nature}\
  }\textbf {\bibinfo {volume} {412}},\ \bibinfo {pages} {313} (\bibinfo {year}
  {2001})}\BibitemShut {NoStop}%
\bibitem [{\citenamefont {Schlederer}\ \emph {et~al.}(2016)\citenamefont
  {Schlederer}, \citenamefont {Krenn}, \citenamefont {Fickler}, \citenamefont
  {Malik},\ and\ \citenamefont {Zeilinger}}]{Schlederer_2016}%
  \BibitemOpen
  \bibfield  {author} {\bibinfo {author} {\bibfnamefont {F.}~\bibnamefont
  {Schlederer}}, \bibinfo {author} {\bibfnamefont {M.}~\bibnamefont {Krenn}},
  \bibinfo {author} {\bibfnamefont {R.}~\bibnamefont {Fickler}}, \bibinfo
  {author} {\bibfnamefont {M.}~\bibnamefont {Malik}}, \ and\ \bibinfo {author}
  {\bibfnamefont {A.}~\bibnamefont {Zeilinger}},\ }\href {\doibase
  10.1088/1367-2630/18/4/043019} {\bibfield  {journal} {\bibinfo  {journal}
  {New Journal of Physics}\ }\textbf {\bibinfo {volume} {18}},\ \bibinfo
  {pages} {043019} (\bibinfo {year} {2016})}\BibitemShut {NoStop}%
\bibitem [{\citenamefont {Huang}\ \emph {et~al.}(2013)\citenamefont {Huang},
  \citenamefont {Ren}, \citenamefont {Yan}, \citenamefont {Ahmed},
  \citenamefont {Yue}, \citenamefont {Bozovich}, \citenamefont {Erkmen},
  \citenamefont {Birnbaum}, \citenamefont {Dolinar}, \citenamefont {Tur},\ and\
  \citenamefont {Willner}}]{Huang:13}%
  \BibitemOpen
  \bibfield  {author} {\bibinfo {author} {\bibfnamefont {H.}~\bibnamefont
  {Huang}}, \bibinfo {author} {\bibfnamefont {Y.}~\bibnamefont {Ren}}, \bibinfo
  {author} {\bibfnamefont {Y.}~\bibnamefont {Yan}}, \bibinfo {author}
  {\bibfnamefont {N.}~\bibnamefont {Ahmed}}, \bibinfo {author} {\bibfnamefont
  {Y.}~\bibnamefont {Yue}}, \bibinfo {author} {\bibfnamefont {A.}~\bibnamefont
  {Bozovich}}, \bibinfo {author} {\bibfnamefont {B.~I.}\ \bibnamefont
  {Erkmen}}, \bibinfo {author} {\bibfnamefont {K.}~\bibnamefont {Birnbaum}},
  \bibinfo {author} {\bibfnamefont {S.}~\bibnamefont {Dolinar}}, \bibinfo
  {author} {\bibfnamefont {M.}~\bibnamefont {Tur}}, \ and\ \bibinfo {author}
  {\bibfnamefont {A.~E.}\ \bibnamefont {Willner}},\ }\href {\doibase
  10.1364/OL.38.002348} {\bibfield  {journal} {\bibinfo  {journal} {Opt.
  Lett.}\ }\textbf {\bibinfo {volume} {38}},\ \bibinfo {pages} {2348} (\bibinfo
  {year} {2013})}\BibitemShut {NoStop}%
\bibitem [{\citenamefont {Sztul}\ and\ \citenamefont
  {Alfano}(2006)}]{Sztul:06}%
  \BibitemOpen
  \bibfield  {author} {\bibinfo {author} {\bibfnamefont {H.~I.}\ \bibnamefont
  {Sztul}}\ and\ \bibinfo {author} {\bibfnamefont {R.~R.}\ \bibnamefont
  {Alfano}},\ }\href {\doibase 10.1364/OL.31.000999} {\bibfield  {journal}
  {\bibinfo  {journal} {Opt. Lett.}\ }\textbf {\bibinfo {volume} {31}},\
  \bibinfo {pages} {999} (\bibinfo {year} {2006})}\BibitemShut {NoStop}%
\bibitem [{\citenamefont {Zhao}\ \emph {et~al.}(2020)\citenamefont {Zhao},
  \citenamefont {Dong}, \citenamefont {Bai},\ and\ \citenamefont
  {Yang}}]{Zhao:20}%
  \BibitemOpen
  \bibfield  {author} {\bibinfo {author} {\bibfnamefont {Q.}~\bibnamefont
  {Zhao}}, \bibinfo {author} {\bibfnamefont {M.}~\bibnamefont {Dong}}, \bibinfo
  {author} {\bibfnamefont {Y.}~\bibnamefont {Bai}}, \ and\ \bibinfo {author}
  {\bibfnamefont {Y.}~\bibnamefont {Yang}},\ }\href {\doibase
  10.1364/PRJ.384925} {\bibfield  {journal} {\bibinfo  {journal} {Photon.
  Res.}\ }\textbf {\bibinfo {volume} {8}},\ \bibinfo {pages} {745} (\bibinfo
  {year} {2020})}\BibitemShut {NoStop}%
\bibitem [{\citenamefont {Zhang}\ \emph {et~al.}(2014)\citenamefont {Zhang},
  \citenamefont {Qi}, \citenamefont {Zhou},\ and\ \citenamefont
  {Chen}}]{PhysRevLett.112.153601}%
  \BibitemOpen
  \bibfield  {author} {\bibinfo {author} {\bibfnamefont {W.}~\bibnamefont
  {Zhang}}, \bibinfo {author} {\bibfnamefont {Q.}~\bibnamefont {Qi}}, \bibinfo
  {author} {\bibfnamefont {J.}~\bibnamefont {Zhou}}, \ and\ \bibinfo {author}
  {\bibfnamefont {L.}~\bibnamefont {Chen}},\ }\href {\doibase
  10.1103/PhysRevLett.112.153601} {\bibfield  {journal} {\bibinfo  {journal}
  {Phys. Rev. Lett.}\ }\textbf {\bibinfo {volume} {112}},\ \bibinfo {pages}
  {153601} (\bibinfo {year} {2014})}\BibitemShut {NoStop}%
\bibitem [{\citenamefont {Dai}\ \emph {et~al.}(2015)\citenamefont {Dai},
  \citenamefont {Gao}, \citenamefont {Zhong}, \citenamefont {Na},\ and\
  \citenamefont {Wang}}]{Dai:15}%
  \BibitemOpen
  \bibfield  {author} {\bibinfo {author} {\bibfnamefont {K.}~\bibnamefont
  {Dai}}, \bibinfo {author} {\bibfnamefont {C.}~\bibnamefont {Gao}}, \bibinfo
  {author} {\bibfnamefont {L.}~\bibnamefont {Zhong}}, \bibinfo {author}
  {\bibfnamefont {Q.}~\bibnamefont {Na}}, \ and\ \bibinfo {author}
  {\bibfnamefont {Q.}~\bibnamefont {Wang}},\ }\href {\doibase
  10.1364/OL.40.000562} {\bibfield  {journal} {\bibinfo  {journal} {Opt.
  Lett.}\ }\textbf {\bibinfo {volume} {40}},\ \bibinfo {pages} {562} (\bibinfo
  {year} {2015})}\BibitemShut {NoStop}%
\bibitem [{\citenamefont {Kotlyar}\ \emph {et~al.}(2017)\citenamefont
  {Kotlyar}, \citenamefont {Kovalev},\ and\ \citenamefont
  {Porfirev}}]{Kotlyar:17}%
  \BibitemOpen
  \bibfield  {author} {\bibinfo {author} {\bibfnamefont {V.~V.}\ \bibnamefont
  {Kotlyar}}, \bibinfo {author} {\bibfnamefont {A.~A.}\ \bibnamefont
  {Kovalev}}, \ and\ \bibinfo {author} {\bibfnamefont {A.~P.}\ \bibnamefont
  {Porfirev}},\ }\href {\doibase 10.1364/AO.56.004095} {\bibfield  {journal}
  {\bibinfo  {journal} {Appl. Opt.}\ }\textbf {\bibinfo {volume} {56}},\
  \bibinfo {pages} {4095} (\bibinfo {year} {2017})}\BibitemShut {NoStop}%
\bibitem [{\citenamefont {Berkhout}\ \emph {et~al.}(2010)\citenamefont
  {Berkhout}, \citenamefont {Lavery}, \citenamefont {Courtial}, \citenamefont
  {Beijersbergen},\ and\ \citenamefont {Padgett}}]{PhysRevLett.105.153601}%
  \BibitemOpen
  \bibfield  {author} {\bibinfo {author} {\bibfnamefont {G.~C.~G.}\
  \bibnamefont {Berkhout}}, \bibinfo {author} {\bibfnamefont {M.~P.~J.}\
  \bibnamefont {Lavery}}, \bibinfo {author} {\bibfnamefont {J.}~\bibnamefont
  {Courtial}}, \bibinfo {author} {\bibfnamefont {M.~W.}\ \bibnamefont
  {Beijersbergen}}, \ and\ \bibinfo {author} {\bibfnamefont {M.~J.}\
  \bibnamefont {Padgett}},\ }\href {\doibase 10.1103/PhysRevLett.105.153601}
  {\bibfield  {journal} {\bibinfo  {journal} {Phys. Rev. Lett.}\ }\textbf
  {\bibinfo {volume} {105}},\ \bibinfo {pages} {153601} (\bibinfo {year}
  {2010})}\BibitemShut {NoStop}%
\bibitem [{\citenamefont {Lavery}\ \emph {et~al.}(2012)\citenamefont {Lavery},
  \citenamefont {Robertson}, \citenamefont {Berkhout}, \citenamefont {Love},
  \citenamefont {Padgett},\ and\ \citenamefont {Courtial}}]{Lavery:12}%
  \BibitemOpen
  \bibfield  {author} {\bibinfo {author} {\bibfnamefont {M.~P.~J.}\
  \bibnamefont {Lavery}}, \bibinfo {author} {\bibfnamefont {D.~J.}\
  \bibnamefont {Robertson}}, \bibinfo {author} {\bibfnamefont {G.~C.~G.}\
  \bibnamefont {Berkhout}}, \bibinfo {author} {\bibfnamefont {G.~D.}\
  \bibnamefont {Love}}, \bibinfo {author} {\bibfnamefont {M.~J.}\ \bibnamefont
  {Padgett}}, \ and\ \bibinfo {author} {\bibfnamefont {J.}~\bibnamefont
  {Courtial}},\ }\href {\doibase 10.1364/OE.20.002110} {\bibfield  {journal}
  {\bibinfo  {journal} {Opt. Express}\ }\textbf {\bibinfo {volume} {20}},\
  \bibinfo {pages} {2110} (\bibinfo {year} {2012})}\BibitemShut {NoStop}%
\bibitem [{\citenamefont {Hickmann}\ \emph {et~al.}(2010)\citenamefont
  {Hickmann}, \citenamefont {Fonseca}, \citenamefont {Soares},\ and\
  \citenamefont {Ch\'avez-Cerda}}]{PhysRevLett.105.053904}%
  \BibitemOpen
  \bibfield  {author} {\bibinfo {author} {\bibfnamefont {J.~M.}\ \bibnamefont
  {Hickmann}}, \bibinfo {author} {\bibfnamefont {E.~J.~S.}\ \bibnamefont
  {Fonseca}}, \bibinfo {author} {\bibfnamefont {W.~C.}\ \bibnamefont {Soares}},
  \ and\ \bibinfo {author} {\bibfnamefont {S.}~\bibnamefont {Ch\'avez-Cerda}},\
  }\href {\doibase 10.1103/PhysRevLett.105.053904} {\bibfield  {journal}
  {\bibinfo  {journal} {Phys. Rev. Lett.}\ }\textbf {\bibinfo {volume} {105}},\
  \bibinfo {pages} {053904} (\bibinfo {year} {2010})}\BibitemShut {NoStop}%
\bibitem [{\citenamefont {Paterson}(2005)}]{PhysRevLett.94.153901}%
  \BibitemOpen
  \bibfield  {author} {\bibinfo {author} {\bibfnamefont {C.}~\bibnamefont
  {Paterson}},\ }\href {\doibase 10.1103/PhysRevLett.94.153901} {\bibfield
  {journal} {\bibinfo  {journal} {Phys. Rev. Lett.}\ }\textbf {\bibinfo
  {volume} {94}},\ \bibinfo {pages} {153901} (\bibinfo {year}
  {2005})}\BibitemShut {NoStop}%
\bibitem [{\citenamefont {Tyler}\ and\ \citenamefont {Boyd}(2009)}]{Tyler:09}%
  \BibitemOpen
  \bibfield  {author} {\bibinfo {author} {\bibfnamefont {G.~A.}\ \bibnamefont
  {Tyler}}\ and\ \bibinfo {author} {\bibfnamefont {R.~W.}\ \bibnamefont
  {Boyd}},\ }\href {\doibase 10.1364/OL.34.000142} {\bibfield  {journal}
  {\bibinfo  {journal} {Opt. Lett.}\ }\textbf {\bibinfo {volume} {34}},\
  \bibinfo {pages} {142} (\bibinfo {year} {2009})}\BibitemShut {NoStop}%
\bibitem [{\citenamefont {Rodenburg}\ \emph {et~al.}(2012)\citenamefont
  {Rodenburg}, \citenamefont {Lavery}, \citenamefont {Malik}, \citenamefont
  {O'Sullivan}, \citenamefont {Mirhosseini}, \citenamefont {Robertson},
  \citenamefont {Padgett},\ and\ \citenamefont {Boyd}}]{Rodenburg:12}%
  \BibitemOpen
  \bibfield  {author} {\bibinfo {author} {\bibfnamefont {B.}~\bibnamefont
  {Rodenburg}}, \bibinfo {author} {\bibfnamefont {M.~P.~J.}\ \bibnamefont
  {Lavery}}, \bibinfo {author} {\bibfnamefont {M.}~\bibnamefont {Malik}},
  \bibinfo {author} {\bibfnamefont {M.~N.}\ \bibnamefont {O'Sullivan}},
  \bibinfo {author} {\bibfnamefont {M.}~\bibnamefont {Mirhosseini}}, \bibinfo
  {author} {\bibfnamefont {D.~J.}\ \bibnamefont {Robertson}}, \bibinfo {author}
  {\bibfnamefont {M.}~\bibnamefont {Padgett}}, \ and\ \bibinfo {author}
  {\bibfnamefont {R.~W.}\ \bibnamefont {Boyd}},\ }\href {\doibase
  10.1364/OL.37.003735} {\bibfield  {journal} {\bibinfo  {journal} {Opt.
  Lett.}\ }\textbf {\bibinfo {volume} {37}},\ \bibinfo {pages} {3735} (\bibinfo
  {year} {2012})}\BibitemShut {NoStop}%
\bibitem [{\citenamefont {Malik}\ \emph {et~al.}(2012)\citenamefont {Malik},
  \citenamefont {O'Sullivan}, \citenamefont {Rodenburg}, \citenamefont
  {Mirhosseini}, \citenamefont {Leach}, \citenamefont {Lavery}, \citenamefont
  {Padgett},\ and\ \citenamefont {Boyd}}]{Malik:12}%
  \BibitemOpen
  \bibfield  {author} {\bibinfo {author} {\bibfnamefont {M.}~\bibnamefont
  {Malik}}, \bibinfo {author} {\bibfnamefont {M.}~\bibnamefont {O'Sullivan}},
  \bibinfo {author} {\bibfnamefont {B.}~\bibnamefont {Rodenburg}}, \bibinfo
  {author} {\bibfnamefont {M.}~\bibnamefont {Mirhosseini}}, \bibinfo {author}
  {\bibfnamefont {J.}~\bibnamefont {Leach}}, \bibinfo {author} {\bibfnamefont
  {M.~P.~J.}\ \bibnamefont {Lavery}}, \bibinfo {author} {\bibfnamefont {M.~J.}\
  \bibnamefont {Padgett}}, \ and\ \bibinfo {author} {\bibfnamefont {R.~W.}\
  \bibnamefont {Boyd}},\ }\href {\doibase 10.1364/OE.20.013195} {\bibfield
  {journal} {\bibinfo  {journal} {Opt. Express}\ }\textbf {\bibinfo {volume}
  {20}},\ \bibinfo {pages} {13195} (\bibinfo {year} {2012})}\BibitemShut
  {NoStop}%
\bibitem [{\citenamefont {Ren}\ \emph {et~al.}(2013)\citenamefont {Ren},
  \citenamefont {Huang}, \citenamefont {Xie}, \citenamefont {Ahmed},
  \citenamefont {Yan}, \citenamefont {Erkmen}, \citenamefont {Chandrasekaran},
  \citenamefont {Lavery}, \citenamefont {Steinhoff}, \citenamefont {Tur},
  \citenamefont {Dolinar}, \citenamefont {Neifeld}, \citenamefont {Padgett},
  \citenamefont {Boyd}, \citenamefont {Shapiro},\ and\ \citenamefont
  {Willner}}]{Ren:13}%
  \BibitemOpen
  \bibfield  {author} {\bibinfo {author} {\bibfnamefont {Y.}~\bibnamefont
  {Ren}}, \bibinfo {author} {\bibfnamefont {H.}~\bibnamefont {Huang}}, \bibinfo
  {author} {\bibfnamefont {G.}~\bibnamefont {Xie}}, \bibinfo {author}
  {\bibfnamefont {N.}~\bibnamefont {Ahmed}}, \bibinfo {author} {\bibfnamefont
  {Y.}~\bibnamefont {Yan}}, \bibinfo {author} {\bibfnamefont {B.~I.}\
  \bibnamefont {Erkmen}}, \bibinfo {author} {\bibfnamefont {N.}~\bibnamefont
  {Chandrasekaran}}, \bibinfo {author} {\bibfnamefont {M.~P.~J.}\ \bibnamefont
  {Lavery}}, \bibinfo {author} {\bibfnamefont {N.~K.}\ \bibnamefont
  {Steinhoff}}, \bibinfo {author} {\bibfnamefont {M.}~\bibnamefont {Tur}},
  \bibinfo {author} {\bibfnamefont {S.}~\bibnamefont {Dolinar}}, \bibinfo
  {author} {\bibfnamefont {M.}~\bibnamefont {Neifeld}}, \bibinfo {author}
  {\bibfnamefont {M.~J.}\ \bibnamefont {Padgett}}, \bibinfo {author}
  {\bibfnamefont {R.~W.}\ \bibnamefont {Boyd}}, \bibinfo {author}
  {\bibfnamefont {J.~H.}\ \bibnamefont {Shapiro}}, \ and\ \bibinfo {author}
  {\bibfnamefont {A.~E.}\ \bibnamefont {Willner}},\ }\href {\doibase
  10.1364/OL.38.004062} {\bibfield  {journal} {\bibinfo  {journal} {Opt.
  Lett.}\ }\textbf {\bibinfo {volume} {38}},\ \bibinfo {pages} {4062} (\bibinfo
  {year} {2013})}\BibitemShut {NoStop}%
\bibitem [{\citenamefont {Aharonov}\ \emph {et~al.}(1988)\citenamefont
  {Aharonov}, \citenamefont {Albert},\ and\ \citenamefont
  {Vaidman}}]{PhysRevLett.60.1351}%
  \BibitemOpen
  \bibfield  {author} {\bibinfo {author} {\bibfnamefont {Y.}~\bibnamefont
  {Aharonov}}, \bibinfo {author} {\bibfnamefont {D.~Z.}\ \bibnamefont
  {Albert}}, \ and\ \bibinfo {author} {\bibfnamefont {L.}~\bibnamefont
  {Vaidman}},\ }\href {\doibase 10.1103/PhysRevLett.60.1351} {\bibfield
  {journal} {\bibinfo  {journal} {Phys. Rev. Lett.}\ }\textbf {\bibinfo
  {volume} {60}},\ \bibinfo {pages} {1351} (\bibinfo {year}
  {1988})}\BibitemShut {NoStop}%
\bibitem [{\citenamefont {Hosten}\ and\ \citenamefont
  {Kwiat}(2008)}]{Hosten787}%
  \BibitemOpen
  \bibfield  {author} {\bibinfo {author} {\bibfnamefont {O.}~\bibnamefont
  {Hosten}}\ and\ \bibinfo {author} {\bibfnamefont {P.}~\bibnamefont {Kwiat}},\
  }\href {\doibase 10.1126/science.1152697} {\bibfield  {journal} {\bibinfo
  {journal} {Science}\ }\textbf {\bibinfo {volume} {319}},\ \bibinfo {pages}
  {787} (\bibinfo {year} {2008})}\BibitemShut {NoStop}%
\bibitem [{\citenamefont {Dixon}\ \emph {et~al.}(2009)\citenamefont {Dixon},
  \citenamefont {Starling}, \citenamefont {Jordan},\ and\ \citenamefont
  {Howell}}]{PhysRevLett.102.173601}%
  \BibitemOpen
  \bibfield  {author} {\bibinfo {author} {\bibfnamefont {P.~B.}\ \bibnamefont
  {Dixon}}, \bibinfo {author} {\bibfnamefont {D.~J.}\ \bibnamefont {Starling}},
  \bibinfo {author} {\bibfnamefont {A.~N.}\ \bibnamefont {Jordan}}, \ and\
  \bibinfo {author} {\bibfnamefont {J.~C.}\ \bibnamefont {Howell}},\ }\href
  {\doibase 10.1103/PhysRevLett.102.173601} {\bibfield  {journal} {\bibinfo
  {journal} {Phys. Rev. Lett.}\ }\textbf {\bibinfo {volume} {102}},\ \bibinfo
  {pages} {173601} (\bibinfo {year} {2009})}\BibitemShut {NoStop}%
\bibitem [{\citenamefont {Lundeen}\ \emph {et~al.}(2011)\citenamefont
  {Lundeen}, \citenamefont {Sutherland}, \citenamefont {Patel}, \citenamefont
  {Stewart},\ and\ \citenamefont {Bamber}}]{lundeen2011direct}%
  \BibitemOpen
  \bibfield  {author} {\bibinfo {author} {\bibfnamefont {J.~S.}\ \bibnamefont
  {Lundeen}}, \bibinfo {author} {\bibfnamefont {B.}~\bibnamefont {Sutherland}},
  \bibinfo {author} {\bibfnamefont {A.}~\bibnamefont {Patel}}, \bibinfo
  {author} {\bibfnamefont {C.}~\bibnamefont {Stewart}}, \ and\ \bibinfo
  {author} {\bibfnamefont {C.}~\bibnamefont {Bamber}},\ }\href {\doibase
  10.1038/nature10120} {\bibfield  {journal} {\bibinfo  {journal} {Nature}\
  }\textbf {\bibinfo {volume} {474}},\ \bibinfo {pages} {188} (\bibinfo {year}
  {2011})}\BibitemShut {NoStop}%
\bibitem [{\citenamefont {Kocsis}\ \emph {et~al.}(2011)\citenamefont {Kocsis},
  \citenamefont {Braverman}, \citenamefont {Ravets}, \citenamefont {Stevens},
  \citenamefont {Mirin}, \citenamefont {Shalm},\ and\ \citenamefont
  {Steinberg}}]{kocsis2011observing}%
  \BibitemOpen
  \bibfield  {author} {\bibinfo {author} {\bibfnamefont {S.}~\bibnamefont
  {Kocsis}}, \bibinfo {author} {\bibfnamefont {B.}~\bibnamefont {Braverman}},
  \bibinfo {author} {\bibfnamefont {S.}~\bibnamefont {Ravets}}, \bibinfo
  {author} {\bibfnamefont {M.~J.}\ \bibnamefont {Stevens}}, \bibinfo {author}
  {\bibfnamefont {R.~P.}\ \bibnamefont {Mirin}}, \bibinfo {author}
  {\bibfnamefont {L.~K.}\ \bibnamefont {Shalm}}, \ and\ \bibinfo {author}
  {\bibfnamefont {A.~M.}\ \bibnamefont {Steinberg}},\ }\href {\doibase
  10.1126/science.1202218} {\bibfield  {journal} {\bibinfo  {journal}
  {Science}\ }\textbf {\bibinfo {volume} {332}},\ \bibinfo {pages} {1170}
  (\bibinfo {year} {2011})}\BibitemShut {NoStop}%
\bibitem [{\citenamefont {Xu}\ \emph {et~al.}(2013)\citenamefont {Xu},
  \citenamefont {Kedem}, \citenamefont {Sun}, \citenamefont {Vaidman},
  \citenamefont {Li},\ and\ \citenamefont {Guo}}]{PhysRevLett.111.033604}%
  \BibitemOpen
  \bibfield  {author} {\bibinfo {author} {\bibfnamefont {X.-Y.}\ \bibnamefont
  {Xu}}, \bibinfo {author} {\bibfnamefont {Y.}~\bibnamefont {Kedem}}, \bibinfo
  {author} {\bibfnamefont {K.}~\bibnamefont {Sun}}, \bibinfo {author}
  {\bibfnamefont {L.}~\bibnamefont {Vaidman}}, \bibinfo {author} {\bibfnamefont
  {C.-F.}\ \bibnamefont {Li}}, \ and\ \bibinfo {author} {\bibfnamefont {G.-C.}\
  \bibnamefont {Guo}},\ }\href {\doibase 10.1103/PhysRevLett.111.033604}
  {\bibfield  {journal} {\bibinfo  {journal} {Phys. Rev. Lett.}\ }\textbf
  {\bibinfo {volume} {111}},\ \bibinfo {pages} {033604} (\bibinfo {year}
  {2013})}\BibitemShut {NoStop}%
\bibitem [{\citenamefont {Topuzoski}\ and\ \citenamefont
  {Janicijevic}(2011)}]{doi:10.1080/09500340.2010.543292}%
  \BibitemOpen
  \bibfield  {author} {\bibinfo {author} {\bibfnamefont {S.}~\bibnamefont
  {Topuzoski}}\ and\ \bibinfo {author} {\bibfnamefont {L.}~\bibnamefont
  {Janicijevic}},\ }\href {\doibase 10.1080/09500340.2010.543292} {\bibfield
  {journal} {\bibinfo  {journal} {Journal of Modern Optics}\ }\textbf {\bibinfo
  {volume} {58}},\ \bibinfo {pages} {138} (\bibinfo {year} {2011})}\BibitemShut
  {NoStop}%
\bibitem [{\citenamefont {Loredo}\ \emph {et~al.}(2009)\citenamefont {Loredo},
  \citenamefont {Ort\'{\i}z}, \citenamefont {Weing\"artner},\ and\
  \citenamefont {De~Zela}}]{PhysRevA.80.012113}%
  \BibitemOpen
  \bibfield  {author} {\bibinfo {author} {\bibfnamefont {J.~C.}\ \bibnamefont
  {Loredo}}, \bibinfo {author} {\bibfnamefont {O.}~\bibnamefont {Ort\'{\i}z}},
  \bibinfo {author} {\bibfnamefont {R.}~\bibnamefont {Weing\"artner}}, \ and\
  \bibinfo {author} {\bibfnamefont {F.}~\bibnamefont {De~Zela}},\ }\href
  {\doibase 10.1103/PhysRevA.80.012113} {\bibfield  {journal} {\bibinfo
  {journal} {Phys. Rev. A}\ }\textbf {\bibinfo {volume} {80}},\ \bibinfo
  {pages} {012113} (\bibinfo {year} {2009})}\BibitemShut {NoStop}%
\bibitem [{\citenamefont {Maga\~na Loaiza}\ \emph {et~al.}(2014)\citenamefont
  {Maga\~na Loaiza}, \citenamefont {Mirhosseini}, \citenamefont {Rodenburg},\
  and\ \citenamefont {Boyd}}]{PhysRevLett.112.200401}%
  \BibitemOpen
  \bibfield  {author} {\bibinfo {author} {\bibfnamefont {O.~S.}\ \bibnamefont
  {Maga\~na Loaiza}}, \bibinfo {author} {\bibfnamefont {M.}~\bibnamefont
  {Mirhosseini}}, \bibinfo {author} {\bibfnamefont {B.}~\bibnamefont
  {Rodenburg}}, \ and\ \bibinfo {author} {\bibfnamefont {R.~W.}\ \bibnamefont
  {Boyd}},\ }\href {\doibase 10.1103/PhysRevLett.112.200401} {\bibfield
  {journal} {\bibinfo  {journal} {Phys. Rev. Lett.}\ }\textbf {\bibinfo
  {volume} {112}},\ \bibinfo {pages} {200401} (\bibinfo {year}
  {2014})}\BibitemShut {NoStop}%
\bibitem [{\citenamefont {Shannon}(1948)}]{6773024}%
  \BibitemOpen
  \bibfield  {author} {\bibinfo {author} {\bibfnamefont {C.~E.}\ \bibnamefont
  {Shannon}},\ }\href {\doibase 10.1002/j.1538-7305.1948.tb01338.x} {\bibfield
  {journal} {\bibinfo  {journal} {The Bell System Technical Journal}\ }\textbf
  {\bibinfo {volume} {27}},\ \bibinfo {pages} {379} (\bibinfo {year}
  {1948})}\BibitemShut {NoStop}%
\bibitem [{\citenamefont {Cover}(1999)}]{cover1999elements}%
  \BibitemOpen
  \bibfield  {author} {\bibinfo {author} {\bibfnamefont {T.~M.}\ \bibnamefont
  {Cover}},\ }\href@noop {} {\emph {\bibinfo {title} {Elements of information
  theory}}}\ (\bibinfo  {publisher} {John Wiley \& Sons},\ \bibinfo {year}
  {1999})\BibitemShut {NoStop}%
\bibitem [{\citenamefont {Lukin}\ \emph {et~al.}(2012)\citenamefont {Lukin},
  \citenamefont {Konyaev},\ and\ \citenamefont {Sennikov}}]{Lukin:12}%
  \BibitemOpen
  \bibfield  {author} {\bibinfo {author} {\bibfnamefont {V.~P.}\ \bibnamefont
  {Lukin}}, \bibinfo {author} {\bibfnamefont {P.~A.}\ \bibnamefont {Konyaev}},
  \ and\ \bibinfo {author} {\bibfnamefont {V.~A.}\ \bibnamefont {Sennikov}},\
  }\href {\doibase 10.1364/AO.51.000C84} {\bibfield  {journal} {\bibinfo
  {journal} {Appl. Opt.}\ }\textbf {\bibinfo {volume} {51}},\ \bibinfo {pages}
  {C84} (\bibinfo {year} {2012})}\BibitemShut {NoStop}%
\bibitem [{\citenamefont {Leach}\ \emph {et~al.}(2002)\citenamefont {Leach},
  \citenamefont {Padgett}, \citenamefont {Barnett}, \citenamefont
  {Franke-Arnold},\ and\ \citenamefont {Courtial}}]{PhysRevLett.88.257901}%
  \BibitemOpen
  \bibfield  {author} {\bibinfo {author} {\bibfnamefont {J.}~\bibnamefont
  {Leach}}, \bibinfo {author} {\bibfnamefont {M.~J.}\ \bibnamefont {Padgett}},
  \bibinfo {author} {\bibfnamefont {S.~M.}\ \bibnamefont {Barnett}}, \bibinfo
  {author} {\bibfnamefont {S.}~\bibnamefont {Franke-Arnold}}, \ and\ \bibinfo
  {author} {\bibfnamefont {J.}~\bibnamefont {Courtial}},\ }\href {\doibase
  10.1103/PhysRevLett.88.257901} {\bibfield  {journal} {\bibinfo  {journal}
  {Phys. Rev. Lett.}\ }\textbf {\bibinfo {volume} {88}},\ \bibinfo {pages}
  {257901} (\bibinfo {year} {2002})}\BibitemShut {NoStop}%
\end{thebibliography}%

\end{document}